\documentclass[lettersize,journal]{IEEEtran}

\usepackage{amsmath,amsfonts,amssymb}
\usepackage{algorithmic}
\usepackage{algorithm}
\usepackage{tabularx}
\usepackage{url}
\usepackage{bm}
\usepackage{makecell}
\usepackage{graphicx}
\usepackage{cite}
\usepackage[caption=false,font=footnotesize,labelfont=footnotesize,textfont=footnotesize]{subfig}
\usepackage{xcolor} 
\begin{document}

\title{Stacked Intelligent Metasurfaces for\\Multi-Modal Semantic Communications}
\author{Guojun Huang, Jiancheng An, \IEEEmembership{Member, IEEE}, Lu Gan, \IEEEmembership{Member, IEEE}, \\ Dusit Niyato, \IEEEmembership{Fellow, IEEE}, Mérouane Debbah, \IEEEmembership{Fellow, IEEE} and Tie Jun Cui, \IEEEmembership{Fellow, IEEE}
\thanks{This work is supported by National Natural Science Foundation of China 62471096.}
\thanks{G. Huang and L. Gan are with the School of Information and Communication Engineering, University of Electronic Science and Technology of China, Chengdu, Sichuan, 611731, China. L. Gan is also with the Yibin Institute of UESTC, Yibin 644000, China (e-mail: guojun$\_$huang2000@sina.com; ganlu@uestc.edu.cn).}
\thanks{J. An is with the School of Electrical and Electronics Engineering, Nanyang Technological University, Singapore 639798 (e-mail: jiancheng\_an@163.com).}
\thanks{D. Niyato is with the College of Computing and Data Science at Nanyang Technological University (NTU), Singapore (e-mail: dniyato@ntu.edu.sg).}
\thanks{M. Debbah is with KU 6G Research Center, Department of Computer and Information Engineering, Khalifa University, Abu Dhabi 127788, UAE (email: merouane.debbah@ku.ac.ae) and also with CentraleSupelec, University Paris-Saclay, 91192 Gif-sur-Yvette, France.}
\thanks{T. J. Cui is with the Institute of Electromagnetic Space, Southeast University, Nanjing, China (e-mail: tjcui@seu.edu.cn).}}
\markboth{Draft}{Draft}
\maketitle

\begin{abstract}
Semantic communication (SemCom) powered by generative artificial intelligence enables highly efficient and reliable information transmission. However, it still necessitates the transmission of substantial amounts of data when dealing with complex scene information. In contrast, the stacked intelligent metasurface (SIM), leveraging wave-domain computing, provides a cost-effective solution for directly imaging complex scenes. Building on this concept, we propose an innovative SIM-aided multi-modal SemCom system. Specifically, an SIM is positioned in front of the transmit antenna for transmitting visual semantic information of complex scenes via imaging on the uniform planar array at the receiver. Furthermore, the simple scene description that contains textual semantic information is transmitted via amplitude-phase modulation over electromagnetic waves. To simultaneously transmit multi-modal information, we optimize the amplitude and phase of meta-atoms in the SIM using a customized gradient descent algorithm. The optimization aims to gradually minimize the mean squared error between the normalized energy distribution on the receiver array and the desired pattern corresponding to the visual semantic information. By combining the textual and visual semantic information, a conditional generative adversarial network is used to recover the complex scene accurately. Extensive numerical results verify the effectiveness of the proposed multi-modal SemCom system in reducing bandwidth overhead as well as the capability of the SIM for imaging the complex scene.
\end{abstract}


\section{Introduction}
\IEEEPARstart{T}{he} next generation of wireless communications seeks to achieve higher transmission rates and lower latency while addressing challenges related to spectrum scarcity \cite{wang2023road, basar2024reconfigurable, gong2023holographic, lin2024efficient}. Semantic communication (SemCom) is envisaged as a novel communication paradigm to solve this challenge by extracting and transmitting semantic information \cite{yang2022semantic, TCCN_2025_Liu_Over}. However, semantic information representation through feature vectors relies on joint optimization, which is time-consuming \cite{Luo2022Semantic}. To address these challenges, generative artificial intelligence (GAI)-assisted SemComs have been developed to enable effective semantic encoding and decoding via separate training. For instance, the authors in \cite{nam2024language} converted an image into a text prompt by using an image-to-text (I2T) encoder -- CLIP \cite{radford2021learning}. The text prompt is subsequently transmitted to the receiver for generating an image through a text-to-image (T2I) tool -- Stable Diffusion \cite{rombach2022high}. In addition to diffusion models, GANs are also widely employed as generative models with faster generation speed. However, uni-modal information may fail to generate accurate sources due to the dynamic nature of generative models \cite{Croitoru_Trans_2023_Diffusion}. To this end, the authors in \cite{du2024generative} combined textual semantic information of the target image with visual semantic information to enhance the source recovery accuracy, at the cost of extra bandwidth overhead. 

Recently, stacked intelligent metasurface (SIM) has emerged as a promising technology to enable complex wave-domain calculations at the speed of light \cite{An2024Stacked, hassan2024efficient, huang2024stacked, TAP_2025_An_Emerging, TCCN_2025_Liu_Multiuser, TWC_2025_Shi_Joint}. Specifically, the authors in \cite{hassan2024efficient} have demonstrated the potential of SIM in achieving radiation image generation. Additionally, the authors in \cite{huang2024stacked} developed an SIM-aided SemCom for image recognition tasks, where the SIM is utilized to convert the transmit signals into a unique beam toward the receiving antenna corresponding to the image class. Nevertheless, the potential of SIM for imaging has not been investigated in a SemCom scenario, albeit with promising benefits for reducing computational requirements and latency.

\begin{figure*}[!t]
\centering
\includegraphics[width=1\textwidth]{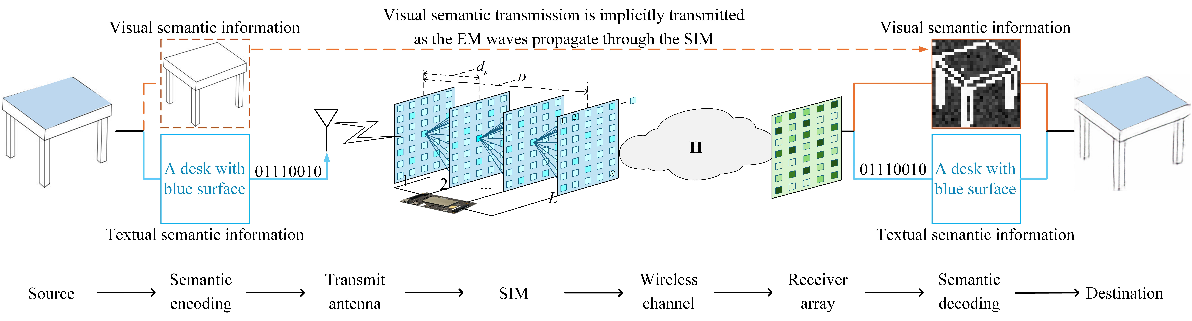}
\caption{An SIM-aided multi-modal SemCom system.}
\label{fig_1}
\end{figure*}

Motivated by this observation, in this letter, we introduce an innovative SIM-aided SemCom system, which aims to achieve the simultaneous transmission of textual and visual semantic information while reducing bandwidth overhead. Specifically, the main contributions are summarized as follows:
\begin{itemize}
 \item We propose a novel SIM-aided multi-modal SemCom system, where an SIM is deployed in front of the transmit antenna to manipulate spatially transmitted electromagnetic (EM) waves, thereby conveying the visual semantic information by directly generating the desired energy distribution on the receiving antenna array, which corresponds to the edge of the complex scene. Simultaneously, the simple scene description that contains textual semantic information is transmitted by a conventional method.
 
 \item To leverage the SIM to generate the desired pattern, we formulate an optimization problem to minimize the mean squared error (MSE) between the normalized energy distribution on the receiving antenna array and the edge of the complex scene and use a customized gradient descent algorithm to optimize the amplitude and phase of the meta-atoms within the SIM.

 \item At the receiver, maximal ratio combining (MRC) is employed to recover the textual semantic information. Additionally, a conditional generative adversarial network (CGAN) is utilized to recover the source by leveraging both textual and visual semantic information. Numerical results demonstrate that the proposed scheme effectively reduces the amount of data transmitted.
\end{itemize}

\section{System Model}
In this section, we introduce the SIM-aided multi-modal SemCom system. Unlike conventional SemCom, an SIM is employed to autonomously generate the desired radiation pattern at the receiving antenna array as EM waves propagate through its layered metasurfaces. Moreover, a CGAN is utilized to recover the image source accurately by combining the textual and visual semantic information.

\subsection{System Model and Task Description}
As shown in Fig. 1, we consider a system consisting of a single antenna, an SIM in front of the antenna, and a receiver array with $M$ antennas. The aim is to accurately recover a complex scene according to the textual and visual semantic information.

\subsubsection{Textual Semantic Information}
The transmit antenna emits the modulated EM waves for carrying the textual semantic information. Specifically, the image source is encoded by a textual semantic encoder into corresponding textual semantics, which is then transmitted through conventional amplitude-phase modulation. Without loss of generality, phase shift keying (PSK) modulation is utilized. Specifically, let $p_{s}$ represent the average transmit power, $\mathcal{O} = \{o_{1}, \ldots, o_{J}\}$ denote the set of constellation points, with $J$ representing the modulation order. As a result, the transmit signal can be represented as $\sqrt{p_{s}}e^{jo_{i}}$, $o_{i} \in \mathcal{O}$.

\subsubsection{Visual Semantic Information}
As shown in Fig. 1, a receiver array with $M$ antennas is configured as a uniform planar array with an antenna spacing of $d_{r}$. We consider that the antennas are arranged in a square configuration. Let $\mathcal{M} = \{1, 2, \ldots, M\}$ denote the set of antenna indices, and the numbers of array antennas per row and per column are denoted by $M_{\mathrm{row}}$ and $M_{\mathrm{col}}$, respectively, satisfying $M = M_{\mathrm{row}}M_{\mathrm{col}}$. The visual semantic information is extracted by the sender through edge detection of the source and then implicitly transmitted by directly imaging on the receiver array.

\subsection{SIM}
The SIM is positioned in front of the transmit antenna to perform complex computations and generate the desired pattern on the receiver array\footnote{In practice, SIM can be integrated with the radome of the base station.} \cite{An2024Stacked}. Let $L$ and $\mathcal{L} = \left \{1, 2, \ldots, L \right \}$ denote the number of layers and the corresponding set, respectively. Let $N$ and $\mathcal{N} = \left \{1, 2, \ldots, N \right \}$ denote the number of meta-atoms on each layer and the corresponding set, respectively. Moreover, let $D$ and $d_s$ denote the thickness of the SIM and the spacing between adjacent layers, respectively, satisfying $d_{s} = D / (L-1)$. Besides, let $z^{l}_{n} = a^{l}_{n} e^{j \phi^{l}_{n}}$ denote the transmission coefficient of the $n$-th meta-atom on the $l$-th metasurface layer \cite{liu2022programmable}, where $0\leq a_{n}^{l}<1$ and $0\leq\phi_{n}^{l}<2\pi$ represent the corresponding amplitude and phase shift, respectively. Let $\mathbf{A}^{l} = \mathrm{diag}\left(\mathbf{a}^{l}\right) = \mathrm{diag} \left(a^{l}_{1}, a^{l}_{2}, \ldots, a^{l}_{N}\right) \in \mathbb{R}^{N \times N}$, $\bm{\Phi}^{l} = \mathrm{diag}\left(e^{j\bm{\phi}^{l}}\right) = \mathrm{diag} \left(e^{j\phi^{l}_{1}}, e^{j\phi^{l}_{2}}, \ldots, e^{j \phi^{l}_{N}}\right) \in \mathbb{C}^{N \times N}$. Thus the transmission coefficient vector and the corresponding matrix that characterizes the response of the $l$-th metasurface layer can be denoted as $\mathbf{z}^{l} = [z^{l}_{1}, z^{l}_{2}, \ldots, z^{l}_{N}]^T \in \mathbb{C}^{N \times 1}$, and $\mathbf{Z}^{l} = \mathrm{diag}\left(\mathbf{z}^{l}\right)\in\mathbb{C}^{N\times N}$, respectively.

Furthermore, the EM propagation matrix from the $(l-1)$-th layer to the $l$-th layer is denoted by $\mathbf{W}^{l} \in \mathbb{C}^{N \times N}$, $l \in \mathcal{L}/\{1\}$. The entry on the $n$-th row and $\check{n}$-th column, i.e., $w^{l}_{n, \check{n}}$ denotes the propagation coefficient from the $\check{n}$-th meta-atom on the $\left(l-1\right)$-th layer to the $n$-th meta-atom on the $l$-th layer. According to Rayleigh-Sommerfeld’s diffraction theory, $w^{l}_{n, \check{n}}$ is defined by \cite{An2024Stacked}
\begin{equation}
\begin{split}
 w^{l}_{n, \check{n}} = \frac{d_{s}S_{a}}{\left({d_{n,\check{n}}^{l}}\right)^{2}}\left(\frac{1}{2\pi d_{n,\check{n}}^{l}}-j\frac{1}{\lambda}\right)e^{j2\pi d_{n,\check{n}}^{l}},
\end{split}
\end{equation}
for $\check{n}, n\in\mathcal{N}$, where $S_{a}$ is the area of each meta-atom, $d^{l}_{n,\check{n}}$ denotes the corresponding propagation distance between the two meta-atoms, and $\lambda$ denotes the radio wavelength. Thus, the EM response of the SIM can be expressed as $\mathbf{B} = \mathbf{Z}^{L} \mathbf{W}^{L} \cdots \mathbf{Z}^{2} \mathbf{W}^{2} \mathbf{Z}^{1} \in \mathbb{C}^{N \times N}$. Additionally, $\mathbf{w}^{1} \in \mathbb{C}^{N \times 1}$ denotes the transmission vector from the transmit antenna to the first metasurface layer \cite{an2023Stacked}, and it can be obtained through replacing $d_{s}$ and $d_{n,\check{n}}^{l}$ in (1) with the horizontal distance from the transmit antenna to the first metasurface, i.e., $d_{t,s}$, and the distance from the transmit antenna to the $n$-th meta-atom on the first layer, i.e., $d^{1}_{n}$, respectively.

\subsection{Signal Model}
The wireless channel $\mathbf{H} = \left[\mathbf{h}_{1}, \mathbf{h}_{2}, \ldots, \mathbf{h}_{M}\right]^T \in \mathbb{C}^{M \times N}$ is assumed to be block fading in this letter, where $\mathbf{h}_{m}^{T} \in \mathbb{C}^{1 \times N}$, $m \in \mathcal{M}$, denotes the channel from the output layer of SIM to the $m$-th antenna at the receiver. Specifically, the Rician fading model \cite{yao2024channel} is considered to model the wireless channel, yielding
\begin{equation}
 \mathbf{H}=\sqrt{\frac{Kq}{1+K}}\mathbf{H}_{\mathrm{LOS}}+\sqrt{\frac{q}{1+K}}\mathbf{H}_{\mathrm{NLOS}}, 
\end{equation}
where $q = C_0d^{-\gamma}$ is the path loss with $C_0$ and $\gamma$ representing the path loss at the reference distance of $1$m and the exponent, respectively, and $d$ is the propagation distance. $K$ is the Rician factor. $\mathbf{H}_{\mathrm{LOS}}$ is the line-of-sight (LOS) component, characterizing the direct path between the SIM and the receiver, whose entry on the $m$-th row and the $n$-column is defined by ${[\mathbf{H}_{\mathrm{LOS}}]}_{m,n} = e^{-j\frac{2\pi}{\lambda}d_{m,n}}$. $d_{m,n}$ denotes the distance from the $n$-th meta-atom on the output layer of the SIM to the $m$-th antenna at the receiving array. $\mathbf{H}_{\mathrm{NLOS}}$ is the non-LOS component, following the complex Gaussian distribution with a zero mean and an identity covariance matrix.

Therefore, the signal received by the $m$-th antenna in the $t$-th time slot can be modeled as
 \begin{equation}
 \label{Eq5}
 y_{m, t} = \sqrt{p_{s}}e^{jo_{t}}\mathbf{h}_{m}^{T}\mathbf{B}\mathbf{w}^{1} + v_{m,t},
 \end{equation}
where $v_{m,t}$ is the additive white Gaussian noise, satisfying $v\sim\mathcal{CN}\left(0, \sigma^{2}_{v}\right)$ with $\sigma^{2}_{v}$ denoting noise power.

\section{Proposed SIM-aided Multi-modal Semantic Information Encoding and Decoding}
In this section, we introduce the detailed semantic encoding and decoding of the SIM-aided multi-modal SemCom system.

\subsection{Semantic Encoder}
The semantic encoding involves two independent components: (i) textual semantic encoding, and (ii) visual semantic encoding.
\subsubsection{Textual Semantic Encoding}
Textual semantic encoding aims to convert image data into a more compact text format for transmission. To this end, existing I2T tools such as CLIP \cite{radford2021learning}, can be employed. Fig. 1 illustrates an example where the transmitter aims to send an image of a desk with blue surface and specific shape to the receiver. The image is processed by a text semantic encoder to extract the key descriptive information, i.e., `A desk with blue surface.' Then, conventional amplitude-phase modulation is utilized to transmit textual semantic information.

\subsubsection{Visual Semantic Encoding}
The aim of visual semantic encoding is to train the SIM to generate a desired radiation pattern on the receiver antenna array by minimizing MSE between the generated radiation pattern and the desired. The desired radiation pattern corresponds to the shape information of the source image which is difficult to characterize using traditional methods \cite{huang2024stacked}.

\subsection{Semantic Decoder}
The semantic decoding involves: (i) textual semantic decoding, and (ii) visual semantic decoding.

\subsubsection{Textual Semantic Decoding}
At the receiver, textual semantic information can be reconstructed by using the conventional multi-antenna detection technique. Specifically, the MRC is utilized to detect the transmit signal, yielding
\begin{equation}
 {\hat{o}_{t}} = \angle \left[\frac{\sum_{m=1}^{M}(\sqrt{p_{s}}\mathbf{h}_{m}^{T}\mathbf{B}\mathbf{w}^{1})^{*}y_{m,t}}{p_{s}||\mathbf{h}_{m}^{T}\mathbf{B}\mathbf{w}^{1}||^{2}}\right].
\end{equation}
The phase ${\hat{o}_{t}}$ is subsequently used for demodulation to recover the textual semantic information.

\subsubsection{Visual Semantic Decoding}
The visual semantic decoder is completed by calculating the normalized received energy distribution pattern. Given a single transmitting antenna and the PSK modulation, the textual semantics do not affect the value of $|y_{m,t}|^2$, and the robustness of visual semantic information transmission can be further enhanced by averaging the results across multiple observation slots.

\subsection{SIM Configuration}
In this section, we elaborate on the SIM configuration to achieve the desired imaging function. Let $\mathbf{Y}^{\mathrm{tg}} \in \mathbb{R}^{M_{\mathrm{row}} \times M_{\mathrm{col}}}$ represent the target radiation pattern, which corresponds to the edge of the source image. Furthermore, let $\mathbf{y}^{\mathrm{tg}} = [y^{\mathrm{tg}}_{1}, y^{\mathrm{tg}}_{2}, \ldots, y^{\mathrm{tg}}_{M}] \in \mathbb{R}^{M \times 1}$ denote the reshaped vector of $\mathbf{Y}^{\mathrm{tg}}$, where $y^{\mathrm{tg}}_{m} \in \{0,1\}, m \in \mathcal{M}$. To enable the SIM to generate the target radiation pattern on the receiver array, we formulate an optimization problem to minimize the MSE between the normalized energy distribution on the receiving antenna array and the edge of the complex scene, yielding 
\begin{equation}
\begin{aligned}
(\mathrm{P}1):\min_{\{\mathbf{A}^{l},\ \mathbf{\Phi}^{l}\}_{l=1}^{L}}& \mathcal{L}=\frac{1}{M}\sum_{m=1}^{M}\left(\zeta|y_{m,t}|^{2}-y^{\mathrm{tg}}_{m}\right)^{2} \\
\mathrm{s.t.}\ & y_{m, t} = \sqrt{p_{s}}e^{jo_{t}}\mathbf{h}_{m}^{T}\mathbf{B}\mathbf{w}^{1} + v_{m,t}, \\
&\mathbf{B}=\mathbf{Z}^{L}\mathbf{W}^{L}\cdots\mathbf{Z}^{2}\mathbf{W}^{2}\mathbf{Z}^{1}, \\
&\mathbf{Z}^{l}=\mathbf{A}^{l}\mathbf{\Phi}^{l}, \\
&0\leq a_{n}^{l}<1,0\leq\phi_{n}^{l}<2\pi,
\end{aligned}
\end{equation}
where $\zeta$ is a scaling factor for normalization \cite{an2023Stacked}, which can be obtained by $\zeta = \sum_{m=1}^{M} y_{m}^{\mathrm{tg}} |y_{m,t}|^{2}/\sum_{m=1}^{M} |y_{m,t}|^4$ at each iteration.

To solve the optimization problem in (5), a gradient descent algorithm is employed to train the transmission coefficients, including the amplitude vectors $\{\mathbf{a}^{1}, \mathbf{a}^{2}, \ldots, \mathbf{a}^{L}\}$ and phase shift vectors $\{\bm{\phi}^{1}, \bm{\phi}^{2}, \ldots, \bm{\phi}^{L}\}$, of the meta-atoms in the SIM for minimizing the MSE $\mathcal{L}$. Moreover, the SIM is initially configured by assuming $a^{l}_{n} \sim \mathcal{U}\left[0, 1\right )$ and $\phi^{l}_{n} \sim \mathcal{U}\left[ 0, 2\pi \right )$, $n \in \mathcal{N}, l \in \mathcal{L}$. 

Specifically, the partial derivatives of $\mathcal{L}$ with respect to $\mathbf{a}^{l}$ and $\bm{\phi}^{l}$ are calculated as follows:
\begin{equation}
 \frac{\partial \mathcal{L}}{\partial{\{\alpha^{l}_{n}, \phi^{l}_{n}\}}} = \frac{2\zeta}{M}\sum_{m=1}^{M}\left(\zeta|y_{m,t}|^{2}-y^{\mathrm{tg}}_{m}\right)\frac{\partial |y_{m,t}|^{2}}{\partial{\{\alpha^{l}_{n}, \phi^{l}_{n}\}}},
\end{equation}
where $\forall l \in \mathcal{L}, \forall n \in \mathcal{N}$. Based on (6), the SIM parameters are updated by
\begin{equation}
 \{\alpha_n^l, \phi_n^l\}\leftarrow\{\alpha_n^l, \phi_n^l\}-\eta\frac{\partial\mathcal{L}}{\partial\{\alpha_n^l, \phi_n^l\}}, \forall n,l,
\end{equation}
where $\eta$ is the learning rate, which is progressively decreased by multiplying $\eta$ with a decay factor $\iota_{f} = 0.8$, when the loss function value does not vary in $50$ consecutive iterations. At each iteration, the SIM transmission coefficients are projected to the feasible set by applying the MinMaxScaler normalization method. Note that the complexity of the SIM configuration method is primarily determined by calculating the partial derivatives. Thus, the computational complexity of the optimization method is $O(ELMN^3)$, where $E$ denotes the number of iterations required.

\begin{table}[t]
\centering
\caption{KEY PARAMETER SETTINGS}
\begin{tabularx}{0.48\textwidth}{|X|l|}
\hline
\multicolumn{2}{|c|}{Physical parameters of SIM} \\ \hline
Thickness of the SIM & $D = 10\lambda$ \\ \hline
Area of each meta-atom & $S_a = \lambda^2$ \\ \hline
Distance between meta-atoms & $r = \lambda$ \\ \hline
Spacing between adjacent layers & $d_s = D/(L-1)$ \\ \hline
\multicolumn{2}{|c|}{System parameters} \\ \hline
Frequency & $f = 28$ GHz \\ \hline
Rician factor & $K = 3$ dB \\ \hline
Number of receive antennas & $M_{\mathrm{row}}$ = 28, $M_{\mathrm{col}}$ = 28 \\ \hline
Spacing between adjacent receive antennas & $d_r = \lambda/2$ \\ \hline
Distance between transmit antenna and SIM & $d_{t,s} = d_s$ \\ \hline
Transmit power & $p_{s} = 40$ dBm \\ \hline
Noise power & $\sigma_{v} = -104$ dBm \\ \hline
Path loss & $C_0 = -35$ dB, $\gamma = 2.8$ \\ \hline
\multicolumn{2}{|c|}{Hyperparameters} \\ \hline
Epoch, learning rate & $E = 2000, \eta = 0.005$ \\ \hline
\end{tabularx}
\label{tab:example}
\end{table}

\section{Simulation Results And Discussion}
This section presents the numerical results to evaluate the effectiveness of the proposed SIM-aided multi-modal SemCom system.

\subsection{Performance versus Key Parameters}
We first evaluate the capability of utilizing SIM to generate the desired radiation pattern, where the parameter setting is shown in Table 1. In Fig. 2, we evaluate the effect of the number $L$ of layers on the generated radiation pattern, where we consider $N = 441$, $d = 5$ m. Specifically, each pixel corresponds to the normalized energy detected by a single antenna, i.e., $\zeta \left | y_{m,t}\right |^{2}$. Note that the MSE decreases as $L$ increases, indicating a closer alignment of the normalized energy distribution with the expected pattern. This improvement is attributed to the enhanced inference capability of the multi-layer diffractive architecture of the SIM, which enables more accurate beam steering for directing energy to desired antennas. Fig. 3 shows the target radiation patterns under different numbers $N$ of meta-atoms on each layer, where $L = 8$ and the other parameters are the same as in Fig. 2. As observed in Fig. 3, the MSE decreases as $N$ increases, since a larger number of meta-atoms would enhance the representational capability of the SIM.

Fig. 4 illustrates the training loss versus epoch under different learning rates, where we set $L = 8$ and maintain the other parameters the same as in Fig. 2. As shown in Fig. 4, convergence is achieved within a few iterations when an appropriate learning rate is selected. Moreover, Fig. 5 verifies the robustness by evaluating the influence of channel estimation errors on the quality of the generated images. Specifically, let $\mathbf{h}^{\prime} = \mathbf{h} + \Delta\mathbf{h}$ represent the practical estimation channel, where $\Delta\mathbf{h}$ represents the channel estimation errors and follows a complex Gaussian distribution, i.e., $\Delta\mathbf{h}\sim\mathcal{CN}\left(0, \sigma^{2}_{h}\mathbf{I_{MN}}\right)$, with $\mathbf{I_{MN}} \in \mathbb{C}^{MN \times MN}$ denoting the identity matrix. In our simulations, $\sigma^{2}_{h}$ is set to $\sigma^{2}_{h} = \beta \Vert \mathbf{h}\Vert^2$, where $\beta$ represents the normalized MSE of channel estimates.

\begin{figure}[t]
\centering
\includegraphics[width=0.48\textwidth]{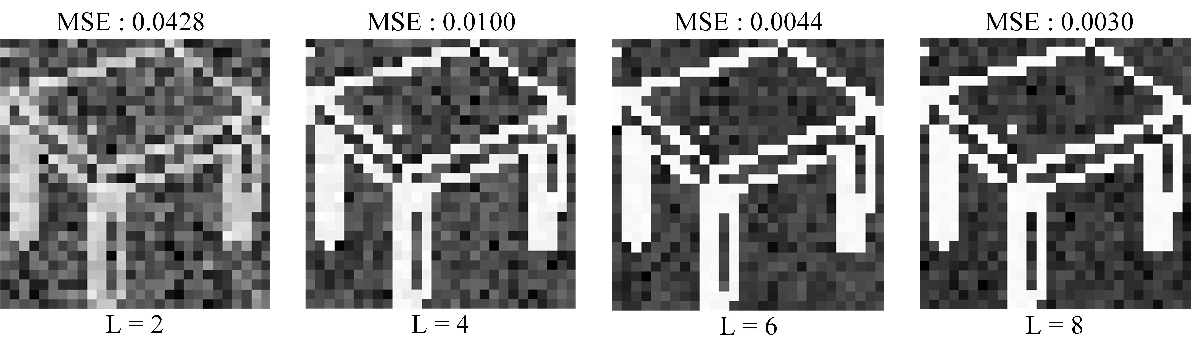}
\caption{The generated radiation pattern under different numbers of metasurface layers.}
\label{fig_2}
\end{figure}

\begin{figure}[t]
\centering
\includegraphics[width=0.48\textwidth]{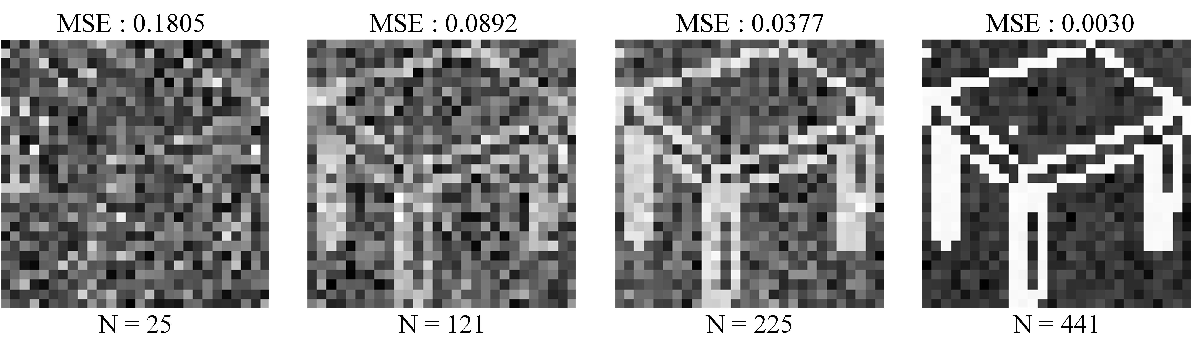}
\caption{The generated radiation pattern under different numbers of meta-atoms on each layer.}
\label{fig_3}
\end{figure}

\subsection{Comparisons with Other Transmission Schemes}
In this subsection, we compare the proposed SemCom system with different GAI-aided transmission schemes. In contrast to GAI-based SemCom systems that utilize pre-trained generative models \cite{qiao2024latency, zhao2024lamosc, cicchetti2024language}, for simplicity, we employ pix2pix \cite{Isola_2017_CVPR} to generate images based on the textual and visual semantic information\footnote{Note that the CGAN is pre-trained based on the shared semantic knowledge base, including a set of images and their corresponding textual and visual semantics. For simplicity, the generated image has $M$ pixels, and StackGAN \cite{Zhang_2017_ICCV} can be incorporated to generate higher-resolution images.}. Moreover, we focus on two key performance metrics: the recovery level of the source image, and the amount of data transmitted. To quantify the latter, the ASCII code is employed for source coding.

\subsubsection{Scheme A}
Scheme A evaluates the performance of conventional SemCom systems. Taking `A desk with blue surface' as an example, to perfectly reconstruct the image at the receiver, it is necessary to provide as detailed textual description or visual semantic feature streams as possible, which accordingly increases the overhead for data transmission. As shown in Fig. 6, even utilizing 792 bits, the receiver still fails to recover the source image to a satisfactory degree, yielding an SSIM value of only 0.6833.
\begin{figure}[t]
\centering
\includegraphics[width=0.45\textwidth]{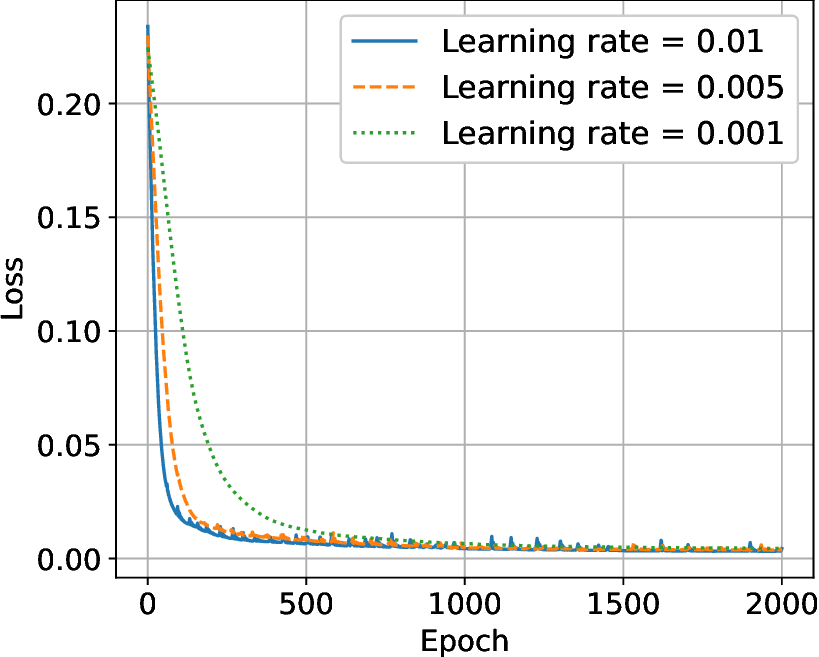}
\caption{Training loss versus epoch.}
\label{Figure4}
\end{figure}

\begin{figure}[t]
\centering
\includegraphics[width=0.45\textwidth]{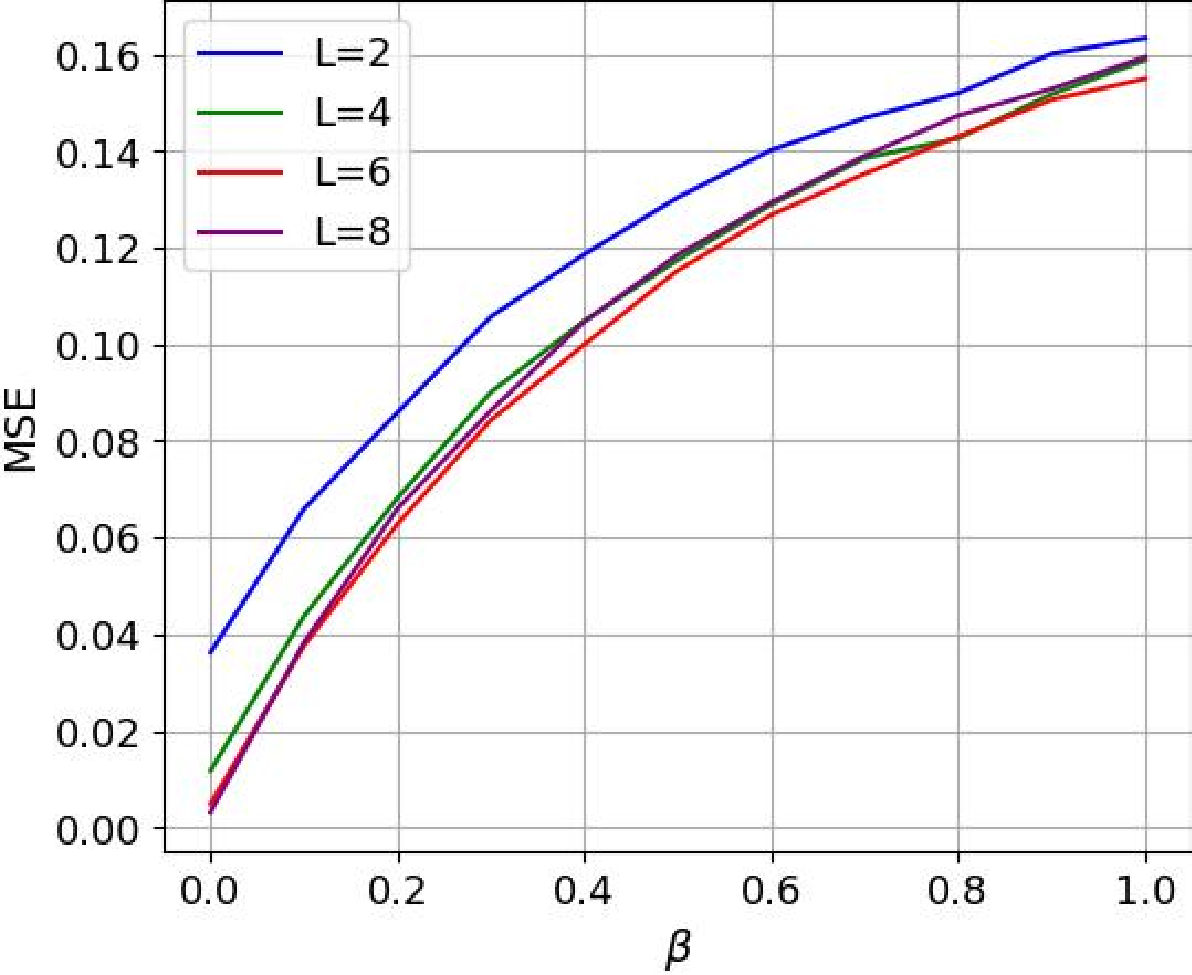}
\caption{MSE versus normalized MSE of channel estimates.}
\label{Figure5}
\end{figure}

\subsubsection{Scheme B}
In Scheme B, we reduce the amount of data transmitted by providing only a simplified description of the source image, which is encoded and transmitted using 192 bits. Since much of the key information is lost, the generated image bears less resemblance to the original one, and the SSIM is only 0.6689.

\subsubsection{Scheme C}
Scheme C directly emits unmodulated EM waves to generate the desired radiation pattern using the SIM. At the receiver, the source image is reconstructed based on the normalized energy distribution pattern. As illustrated in Fig. 6, based on the visual semantic information, the profile of the source image can be recovered with zero overhead, achieving an SSIM of 0.8194. While it exhibits a structural resemblance to the source image, color information is missing.

\subsubsection{Scheme D}
Scheme D evaluates the performance of the proposed SIM-aided SemCom system. By combining the strengths of Scheme B and C, Scheme D enables the simultaneous transmission of textual and visual semantic information by modulating only a small amount of information onto EM waves. Nonetheless, the receiver achieves precise source reconstruction by leveraging multi-modal information, achieving a satisfactory SSIM of 0.8855. Remarkably, the transmission of visual semantic information occurs naturally as EM waves propagate through the SIM, without any bandwidth consumption.

\begin{figure}[t]
\centering
\includegraphics[width=0.48\textwidth]{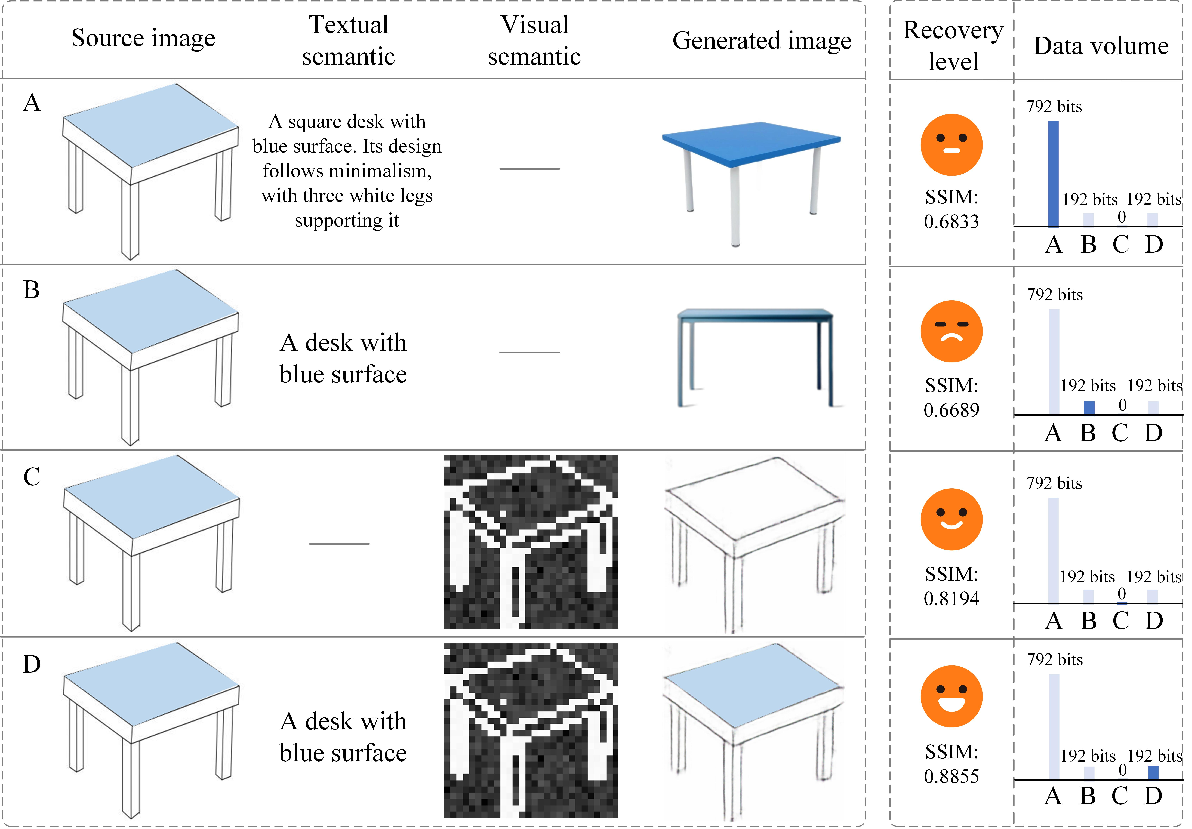}
\caption{Performance comparison of different GAI-aided SemCom schemes, where $N = 8, L = 6$, and $d = 5$ m.}
\label{fig_5}
\end{figure}

\section{Conclusion}
We have presented an SIM-aided multi-modal SemCom system to transmit complex scene information with reduced overhead. Specifically, an SIM in front of the transmit antenna has been optimized to steer the modulation signals toward diverse receive antennas to produce the expected radiation pattern on the receiver array. Furthermore, the classic MRC has been utilized to recover the textual semantic information. By employing multi-modal semantic information, a CGAN has been used to recover the source image. Extensive simulation results have verified the effectiveness of the proposed multi-modal SemCom system in reducing bandwidth overhead. Note that this letter only explores the bandwidth reduction for transmitting visual semantics, while the interplay between SIM-aided SemCom and other types of data transmission remains a topic for further investigation.

\bibliographystyle{IEEEtran}
\bibliography{IEEEabrv,Ref}

\begin{thebibliography}{10}
\providecommand{\url}[1]{#1}
\csname url@samestyle\endcsname
\providecommand{\newblock}{\relax}
\providecommand{\bibinfo}[2]{#2}
\providecommand{\BIBentrySTDinterwordspacing}{\spaceskip=0pt\relax}
\providecommand{\BIBentryALTinterwordstretchfactor}{4}
\providecommand{\BIBentryALTinterwordspacing}{\spaceskip=\fontdimen2\font plus
\BIBentryALTinterwordstretchfactor\fontdimen3\font minus \fontdimen4\font\relax}
\providecommand{\BIBforeignlanguage}[2]{{%
\expandafter\ifx\csname l@#1\endcsname\relax
\typeout{** WARNING: IEEEtran.bst: No hyphenation pattern has been}%
\typeout{** loaded for the language `#1'. Using the pattern for}%
\typeout{** the default language instead.}%
\else
\language=\csname l@#1\endcsname
\fi
#2}}
\providecommand{\BIBdecl}{\relax}
\BIBdecl

\bibitem{wang2023road}
C.-X. Wang, X.~You, X.~Gao, X.~Zhu, Z.~Li, C.~Zhang, H.~Wang, Y.~Huang, Y.~Chen, H.~Haas \emph{et~al.}, ``On the road to {6G}: Visions, requirements, key technologies and testbeds,'' \emph{IEEE Commun. Surveys Tuts.}, vol.~25, no.~2, pp. 905--974, Feb. 2023.

\bibitem{basar2024reconfigurable}
E.~Basar, G.~C. Alexandropoulos, Y.~Liu, Q.~Wu, S.~Jin, C.~Yuen, O.~A. Dobre, and R.~Schober, ``Reconfigurable intelligent surfaces for {6G}: Emerging hardware architectures, applications, and open challenges,'' \emph{IEEE Vehicular Technology Magazine}, Jul. 2024.

\bibitem{gong2023holographic}
T.~Gong, P.~Gavriilidis, R.~Ji, C.~Huang, G.~C. Alexandropoulos, L.~Wei, Z.~Zhang, M.~Debbah, H.~V. Poor, and C.~Yuen, ``Holographic {MIMO} communications: Theoretical foundations, enabling technologies, and future directions,'' \emph{IEEE Communications Surveys \& Tutorials}, vol.~26, no.~1, pp. 196--257, Aug. 2023.

\bibitem{lin2024efficient}
Z.~Lin, G.~Zhu, Y.~Deng, X.~Chen, and Y.~Gao, ``Efficient parallel split learning over resource-constrained wireless edge networks,'' \emph{IEEE Trans. Mobile Comput.}, vol.~23, no.~10, pp. 9224--9239, Jan. 2024.

\bibitem{yang2022semantic}
W.~Yang, H.~Du, Z.~Q. Liew, W.~Y.~B. Lim, Z.~Xiong, D.~Niyato, X.~Chi, X.~S. Shen, and C.~Miao, ``Semantic communications for future internet: Fundamentals, applications, and challenges,'' \emph{IEEE Commun. Surveys Tuts.}, vol.~25, no.~1, pp. 213--250, Nov. 2022.

\bibitem{TCCN_2025_Liu_Over}
M.~Liu \emph{et~al.}, ``Over-the-air ode-inspired neural network for dual task-oriented semantic communications,'' \emph{IEEE Trans. Cognitive Commun. Netw.}, pp. 1--1, 2025, Early Access.

\bibitem{Luo2022Semantic}
X.~Luo, H.-H. Chen, and Q.~Guo, ``Semantic communications: Overview, open issues, and future research directions,'' \emph{IEEE Wireless Commun.}, vol.~29, no.~1, pp. 210--219, Jan. 2022.

\bibitem{nam2024language}
H.~Nam, J.~Park, J.~Choi, M.~Bennis, and S.-L. Kim, ``Language-oriented communication with semantic coding and knowledge distillation for text-to-image generation,'' in \emph{ICASSP 2024-2024 IEEE International Conference on Acoustics, Speech and Signal Processing (ICASSP)}.\hskip 1em plus 0.5em minus 0.4em\relax IEEE, 2024, pp. 13\,506--13\,510.

\bibitem{radford2021learning}
A.~Radford, J.~W. Kim, C.~Hallacy, A.~Ramesh, G.~Goh, S.~Agarwal, G.~Sastry, A.~Askell, P.~Mishkin, J.~Clark \emph{et~al.}, ``Learning transferable visual models from natural language supervision,'' in \emph{International conference on machine learning}.\hskip 1em plus 0.5em minus 0.4em\relax PMLR, Jul. 2021, pp. 8748--8763.

\bibitem{rombach2022high}
R.~Rombach, A.~Blattmann, D.~Lorenz, P.~Esser, and B.~Ommer, ``High-resolution image synthesis with latent diffusion models,'' in \emph{Proceedings of the IEEE/CVF conference on computer vision and pattern recognition}, Jun. 2022, pp. 10\,684--10\,695.

\bibitem{Croitoru_Trans_2023_Diffusion}
F.-A. Croitoru, V.~Hondru, R.~T. Ionescu, and M.~Shah, ``Diffusion models in vision: A survey,'' \emph{IEEE Trans. Pattern Anal. Mach. Intell.}, vol.~45, no.~9, pp. 10\,850--10\,869, Mar. 2023.

\bibitem{du2024generative}
H.~Du, G.~Liu, D.~Niyato, J.~Zhang, J.~Kang, Z.~Xiong, B.~Ai, and D.~I. Kim, ``Generative {AI}-aided joint training-free secure semantic communications via multi-modal prompts,'' in \emph{ICASSP 2024-2024 IEEE International Conference on Acoustics, Speech and Signal Processing (ICASSP)}.\hskip 1em plus 0.5em minus 0.4em\relax IEEE, Mar. 2024, pp. 12\,896--12\,900.

\bibitem{An2024Stacked}
J.~An, C.~Yuen, C.~Xu, H.~Li, D.~W.~K. Ng, M.~D. Renzo, M.~Debbah, and L.~Hanzo, ``Stacked intelligent metasurface-aided {MIMO} transceiver design,'' \emph{IEEE Wireless Commun.}, pp. 1--9, 2024, {E}arly Access.

\bibitem{hassan2024efficient}
N.~U. Hassan, J.~An, M.~Di~Renzo, and M.~Debbah, ``Efficient beamforming and radiation pattern control using stacked intelligent metasurfaces,'' \emph{IEEE Open J. Commun. Soc.}, vol.~5, pp. 599--611, Jan. 2024.

\bibitem{huang2024stacked}
G.~Huang, J.~An, Z.~Yang, L.~Gan, M.~Bennis, and M.~Debbah, ``Stacked intelligent metasurfaces for task-oriented semantic communications,'' \emph{IEEE Wireless Commun. Lett.}, pp. 1--1, Nov. 2024.

\bibitem{TAP_2025_An_Emerging}
J.~An, M.~Debbah, T.~J. Cui, Z.~N. Chen, and C.~Yuen, ``Emerging technologies in intelligent metasurfaces: Shaping the future of wireless communications,'' \emph{IEEE Trans. Antennas Propag.}, pp. 1--1, 2025, Early Access.

\bibitem{TCCN_2025_Liu_Multiuser}
H.~Liu \emph{et~al.}, ``Multi-user {MISO} with stacked intelligent metasurfaces: A {DRL}-based sum-rate optimization approach,'' \emph{IEEE Trans. Cognitive Commun. Netw.}, pp. 1--1, 2025, Early Access.

\bibitem{TWC_2025_Shi_Joint}
E.~Shi \emph{et~al.}, ``Joint {AP-UE} association and precoding for {SIM}-aided cell-free massive {MIMO} systems,'' \emph{IEEE Trans. Wireless Commun.}, vol.~24, no.~6, pp. 5352--5367, Jun. 2025.

\bibitem{liu2022programmable}
C.~Liu, Q.~Ma, Z.~J. Luo, Q.~R. Hong, Q.~Xiao, H.~C. Zhang, L.~Miao, W.~M. Yu, Q.~Cheng, L.~Li \emph{et~al.}, ``A programmable diffractive deep neural network based on a digital-coding metasurface array,'' \emph{Nat. Electron.}, vol.~5, no.~2, pp. 113--122, Feb. 2022.

\bibitem{an2023Stacked}
J.~An, C.~Xu, D.~W.~K. Ng, G.~C. Alexandropoulos, C.~Huang, C.~Yuen, and L.~Hanzo, ``Stacked intelligent metasurfaces for efficient holographic {MIMO} communications in {6G},'' \emph{IEEE J. Sel. Areas Commun.}, vol.~41, no.~8, pp. 2380--2396, Jun. 2023.

\bibitem{yao2024channel}
X.~Yao, J.~An, L.~Gan, M.~Di~Renzo, and C.~Yuen, ``Channel estimation for stacked intelligent metasurface-assisted wireless networks,'' \emph{IEEE Wireless Commun. Lett.}, vol.~13, no.~5, pp. 1349--1353, May. 2024.

\bibitem{qiao2024latency}
L.~Qiao, M.~B. Mashhadi, Z.~Gao, C.~H. Foh, P.~Xiao, and M.~Bennis, ``Latency-aware generative semantic communications with pre-trained diffusion models,'' \emph{IEEE Wireless Commun. Lett.}, Jul. 2024.

\bibitem{zhao2024lamosc}
Y.~Zhao, Y.~Yue, S.~Hou, B.~Cheng, and Y.~Huang, ``Lamosc: Large language model-driven semantic communication system for visual transmission,'' \emph{IEEE Trans. Cogn. Commun. Netw.}, May. 2024.

\bibitem{cicchetti2024language}
G.~Cicchetti, E.~Grassucci, J.~Park, J.~Choi, S.~Barbarossa, and D.~Comminiello, ``Language-oriented semantic latent representation for image transmission,'' in \emph{2024 IEEE 34th International Workshop on Machine Learning for Signal Processing (MLSP)}.\hskip 1em plus 0.5em minus 0.4em\relax IEEE, Nov. 2024, pp. 1--6.

\bibitem{Isola_2017_CVPR}
P.~Isola, J.-Y. Zhu, T.~Zhou, and A.~A. Efros, ``Image-to-image translation with conditional adversarial networks,'' in \emph{Proceedings of the IEEE Conference on Computer Vision and Pattern Recognition (CVPR)}, Jul. 2017.

\bibitem{Zhang_2017_ICCV}
H.~Zhang, T.~Xu, H.~Li, S.~Zhang, X.~Wang, X.~Huang, and D.~N. Metaxas, ``Stackgan: Text to photo-realistic image synthesis with stacked generative adversarial networks,'' in \emph{Proceedings of the IEEE International Conference on Computer Vision (ICCV)}, Oct. 2017.

\end{thebibliography}
\end{document}